\documentclass[aps,prl,prb,twocolumn,showpacs,superscriptaddress,groupedaddress]{revtex4-1}
\usepackage{graphicx}
\usepackage{physics}
\usepackage{amssymb}
\usepackage{amsmath}
\usepackage{latexsym}
\usepackage{ulem}
\usepackage{color}
\usepackage{dcolumn}
\usepackage{subfigure}
\usepackage[T1]{fontenc}
\usepackage[utf8]{inputenc}
\usepackage[english,french]{babel}
\usepackage[colorlinks=true,linkcolor=blue,citecolor=blue]{hyperref} %
\usepackage{bm}        
\usepackage{amssymb}   
\usepackage{booktabs}

\usepackage[capitalize]{cleveref}

\usepackage{lipsum}
\usepackage[dvipsnames]{xcolor}

\begin{document}

\graphicspath{{figures/}}

\definecolor{airforceblue}{rgb}{0.36, 0.54, 0.66}

\title{Strengthening of the superconductivity by real space decimation \\ of the flat band states}

\author{M. Thumin}%
 \email{maxime.thumin@neel.cnrs.fr}
 \author{G. Bouzerar}%
 \email{georges.bouzerar@neel.cnrs.fr}
\affiliation{Université Grenoble Alpes, CNRS, Institut NEEL, F-38042 Grenoble, France}%
\date{\today}

\begin{abstract}
In contrast to standard BCS superconductivity, that in flat bands (FBs) possesses an interesting degree of freedom that enables the control of the superfluid weight (SFW), referred to as the quantum metric (QM). In the present work, we consider the stub lattice and study the impact of the dilution of FB eigenstates on superconductivity. Among the most remarkable results, it is revealed that the SFW can be boosted by the decimation of the FB eigenstates. In addition, it is shown that the widely used uniform pairing hypothesis systematically predicts the suppression of the SFW, appears misleading and qualitatively incorrect. 
With the great progress in nanotechnologies, we believe that our findings could be realised and tested experimentally in covalent organic frameworks or in decorated structures in which defects/vacancies/ad-atoms are created/deposited in a controlled manner and even in multilayered structures with intercalated atoms.
\end{abstract} 

\maketitle


\section{Introduction}
It is well established that electronic Bloch states with infinitely narrow bands, known as flat bands (FBs), provide a unique platform for studying strongly correlated phases of matter 
\cite{Leykam_review, BCS_FB, Tasaki_magnetism, FB_magnetism_Bouzerar, Frac_Hall_1, Frac_Hall_2, Frac_Hall_3, Cao2018_1, Efetov}. One of the most striking phenomenon is the possibility of superconductivity (SC) 
in dispersion-less bands. This intriguing unconventional form of SC, possesses unusual scaling laws and opens promising avenues towards room-temperature SC. Indeed, the critical temperature scales linearly with the strength of the effective attractive electron-electron interaction $T_c \propto |U|$ \cite{BCS_FB,Volovik_T_linear, Volovik_T_linear_Flat_bands_in_topological_media}.
Hence, SC in FBs should lead to significantly higher critical temperature than that in conventional BCS superconductors which predicts $T_c^{BCS} \propto t \, e^{-1/\rho(E_F)|U|}$ \cite{BCS} where $\rho(E_F)$ denotes the density of states at the Fermi level.
Furthermore, the unusual nature of the of superconductivity in FBs arises from its connection with quantum geometry \cite{Peotta_Nature,Peotta_Lieb, Iskin}, and more precisely with the quantum metric (QM) \cite{Provost_metric,Berry_5_years,Resta}. More specifically, it has been predicted that the superfluid weight (SFW) $D_s$, a key quantity to calculate the critical temperature \cite{Jump_Ds}, can be written $D_s \propto |U| \langle g \rangle n^{-1}_{S}$ where $\langle g \rangle$ is the average value of the QM and $n_{S}$ is the number of orbitals for which the FB weight is non zero \cite{Batrouni_sawtooth,Torma_revisiting}. We remark, that this prediction is valid when $|U|$ is smaller than the one-particle gap.
We recall that $\langle g \rangle$ has a geometric interpretation since it provides a measure of the square of a typical lengthscale associated to the FB eigenstates. Indeed, the diagonal components of the QM can be expressed as the square of the standard deviation of the position operator $\hat{x}_\mu$ relatively to a FB state $|\psi_k^{FB}\rangle$, namely it can be shown that $g_{\mu\mu}(k) = \langle  \psi_k^{FB}| \hat{x}_\mu^2 - \langle \hat{x}_\mu \rangle^2 |\psi_k^{FB}\rangle$
\cite{Marzari_spread_PRB, Marzari_sprea_revew}.
It is important for what follows to emphasize that the QM is not uniquely defined, it depends on the relative position of the orbitals inside the unit cell. However, in the expression of the SFW, $\langle g \rangle$, means minimal value of the QM \cite{Torma_revisiting} which in general, corresponds to the highly symmetric positions of the orbitals.
 
\begin{figure}[h!]
\centering
\includegraphics[scale=0.23]{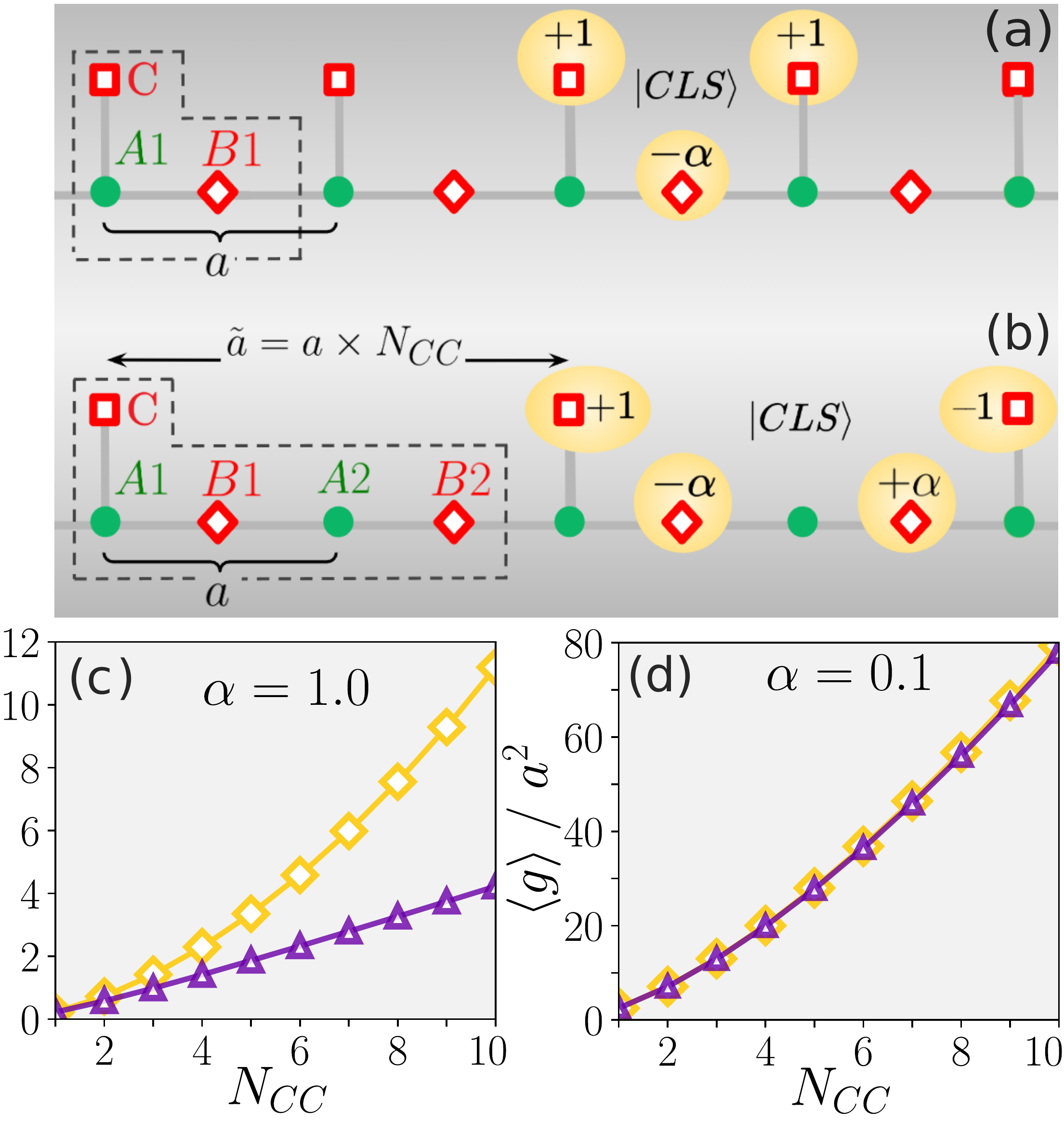} 
\caption{\textbf{(a)} Representation of the stub lattice (reference) and its compact localized state (CLS). \textbf{(b)} The diluted stub lattice with $N_{CC}=2$ and its CLS. In (a) and (b), green (resp. red) symbols correspond to orbitals in sublattice $\mathcal{A}$ (resp. $\mathcal{BC}$). The grey dashed lines depict the unit cells.
\textbf{(c)} and \textbf{(d)} display the QM $\langle g \rangle$ as a function of $N_{CC}$ for $\alpha=1.0$ and $\alpha=0.1$. The diamonds correspond to the orbital positions in the unit cell as depicted in (a),(b) and the triangles to the minimal value of the QM (see main text). 
}
\label{Fig. 1}
\end{figure}

\begin{figure*}[ht]
    \centering
\includegraphics[scale=0.58]{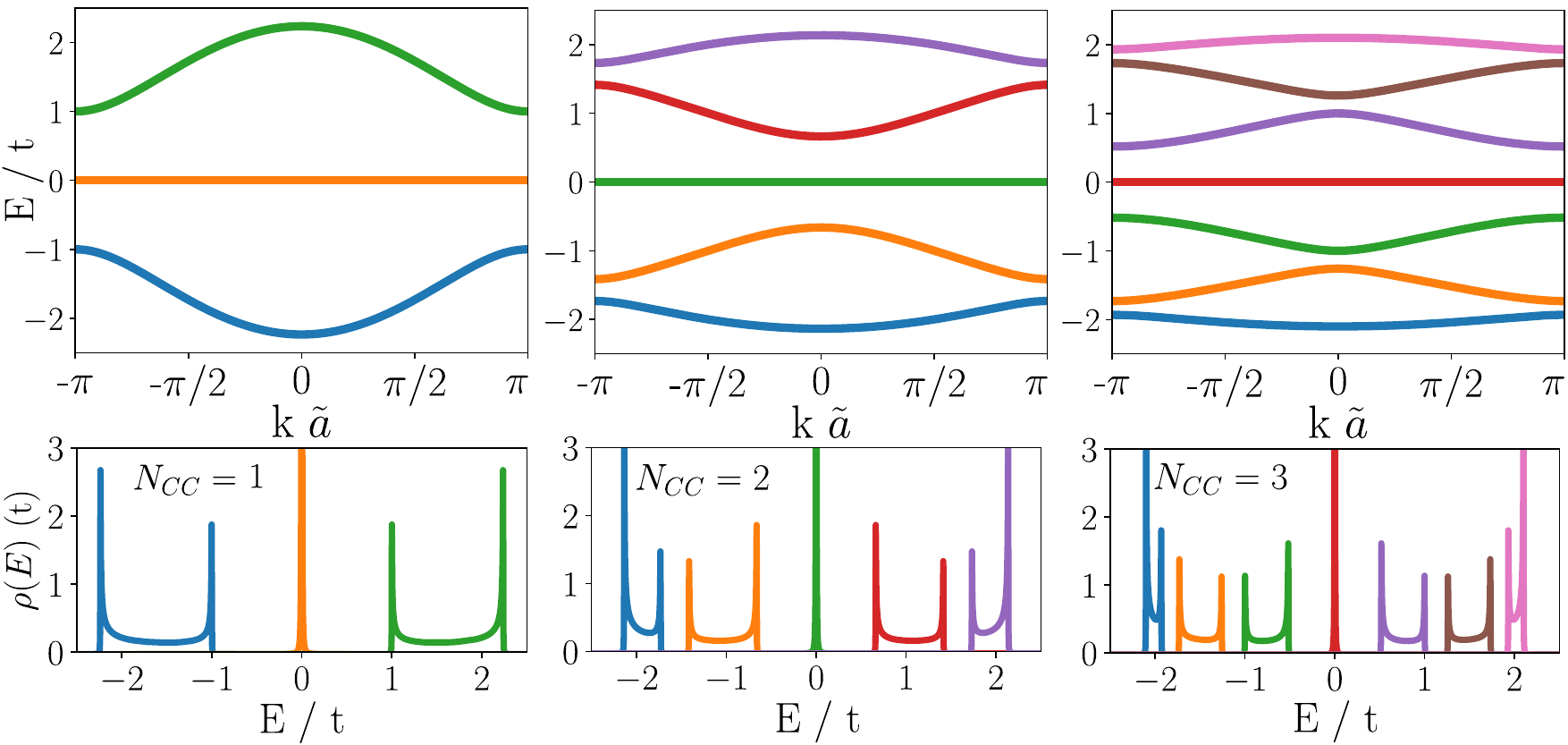} 
     \caption{Single-particle spectrum (top) and corresponding density of states (bottom) for $\alpha = 1.0$, and $N_{CC}= 1, 2,$ and $3$ respectively. $\tilde{a}=N_{CC} \cdot a$ is the super-lattice parameter. In the DOS panels the colors correspond to that of the bands depicted in the dispersion curves.  
     } 
 \label{Fig. 2}
\end{figure*} 

In real systems, whereas $|U|$ is essentially fixed by the chemical structure of the compound, the QM still offers a tunable degree of freedom to increase $D_s$ and ultimately $T_c$. To reach the "holy grail" of room-temperature SC with FB systems, the search for a boosting mechanism of the QM is a versatile strategy. Recently, it has been shown that the QM can be significantly enhanced by the introduction of vacancies that leads to the spreading out the FB eigenstates \cite{giant_boost}. 
Hence, a natural question arises: \textit{Does the SFW increase systematically when the FB states are spreading out ?}
or equivalently \textit{Does the dilution of the FB eigenstates reinforce the superconducting phase ?}

Here, we propose to address precisely this issue by considering the diluted stub lattice. The conventional stub lattice is illustrated in Fig$\,$\ref{Fig. 1}(a), it consists in three orbitals ($A,B$, and $C$) per unit cell with a FB located at $E=0$. Its compact localized eigenstates (CLS) span two cells. By removing every second $C$-orbital along the chain, the CLS spreads up as depicted in Fig$\,$\ref{Fig. 1}(b) and now spans three cells of the conventional lattice. We can continue the process and remove two out of three $C$-orbitals, we thus generate the next generation of dilute stub chain, and so on. The decimation procedure is clear. In what follows, it is convenient to characterize the dilute stub chain by introducing the integer $N_{CC}$ where $\tilde{a} = N_{CC} \times a$ is the distance between two successive $C$-orbitals along the chain. The detailed study of the conventional stub chain ($N_{CC}=1$) has been realized in Ref.~\cite{Thumin_PRB_2023}. \\
Notice that, it is important not to confuse the spread of CLS states and that of the QM. Nevertheless, a larger CLS implies a larger $\langle g \rangle$. We should also emphasize that in most of the studies found in the literature $n_{S}$ is fixed and hence irrelevant. In contrast, here, the procedure of decimation affects simultaneously $\langle g \rangle$ and $n_{S}$. 


\begin{figure*}
    \centering
\includegraphics[scale=0.6]{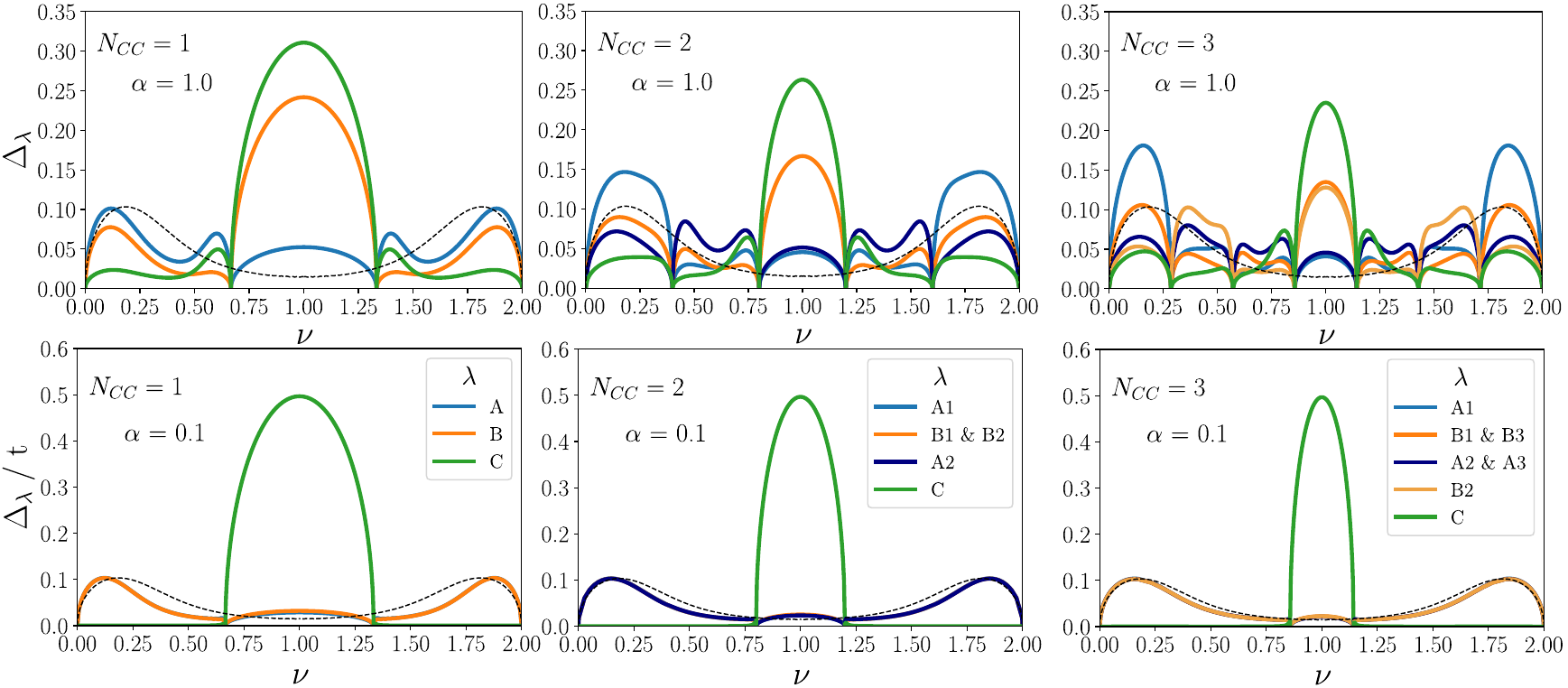}  
     \caption{Pairings $\Delta_\lambda$, as a function of the filling factor $\nu$. 
     The upper panel corresponds to $\alpha=1.0$, and $N_{CC}=1, 2$ and $3$ (left to right). The lower panel corresponds to $\alpha=0.1$ and the same values of $N_{CC}$. The interaction strength is fixed $|U|/t=1$. The black dashed line is the pairing in the one dimensional chain ($\alpha = 0$).}
 \label{Fig. 3}
\end{figure*}

\section{Model and methods}
To address the SC in the depleted stub lattice, the electrons are described by the attractive Hubbard model (AHM) that reads,
\begin{equation}
\begin{split}
    \hat{H} = \sum_{\langle i\lambda,j\eta \rangle, \sigma} t^{\lambda\eta}_{ij} \; \hat{c}_{i\lambda, \sigma}^{\dagger} \hat{c}_{j \eta, \sigma} - \mu \hat{N} - |U| \sum_{i\lambda} \hat{n}_{i\lambda\uparrow}\hat{n}_{i\lambda\downarrow}
    \label{H_exact}
\end{split}
,
\end{equation}
where $\hat{c}_{i \lambda \sigma}^{\dagger}$ creates an electron of spin $\sigma$ at site $\textbf{r}_{i\lambda}$, $i$ being the cell index and $\lambda = A_1, B_1,...,A_{N_{CC}}, B_{N_{CC}}$ and $C$ the $2N_{CC} + 1$ orbitals in the unit cell. For practical reason and because the system is bipartite, we define two families of orbitals (or sublattices) $\mathcal{A}=\{A_1,...,A_{N_{CC}} \}$ and $\mathcal{BC}=\{B_1,...,B_{N_{CC}},C \}$.
Sums run over the lattice, $\langle i\lambda,j \eta\rangle$ refers to nearest-neighbor pairs for which the hopping integral $t^{\lambda\eta}_{ij}$ is $t$ along the chain, and $\alpha t$ for the out-of-chain hopping. 
The parameter $\alpha$ allows the tuning of the single particle gap and of the QM. The particle number operator is $\hat{N}=\sum_{i\lambda,\sigma} \hat{n}_{i\lambda,\sigma}$, and $\mu$ is the chemical potential. Finally, $|U|$ is the strength of the on-site attractive electron-electron interaction. 

In the specific case of FB superconductivity, the mean-field Bogoliubov-de-Gennes (BdG) approach has been shown to be a reliable and remarkably accurate to calculate both the occupations, the pairings, the one-particle correlation functions and the transport properties regardless the spatial dimension \cite{Peotta_Lieb, Batrouni_CuO2,Batrouni_sawtooth, Batrouni_Designer_Flat_Bands,Hofmann_PRB,Chi_QMC,Peri_PRL,Arbeitman_PRL, Thumin_coherence_length}. Furthermore, in bipartite system with gapped FBs, the BCS ground state has been shown to be the exact ground state of the AHM \eqref{H_exact} when the interaction strength is smaller than the gap \cite{Peotta_Nature, Peotta_Lieb}. These conditions being fulfilled here, one can safely and confidently study the SC in FBs within the BdG theory. \\
We briefly recall that the BdG approach consists in decoupling the Hubbard interaction term as follows,
\begin{equation}
\begin{split}
\hat{n}_{i\lambda,\uparrow}\hat{n}_{i\lambda,\downarrow} \overset{BdG}{\simeq}
&\langle\hat{n}_{i\lambda,\downarrow}\rangle  \hat{n}_{i\lambda,\uparrow} +\langle\hat{n}_{i\lambda,\uparrow}\rangle  \hat{n}_{i\lambda,\downarrow} \\
- \; &\,\frac{\Delta_{i\lambda}}{|U|} \hat{c}^{\dagger}_{i\lambda,\uparrow}\hat{c}^{\dagger}_{i\lambda,\downarrow} 
- \frac{\Delta_{i\lambda}^*}{|U|} \hat{c}_{i\lambda,\downarrow}\hat{c}_{i\lambda,\uparrow}  \\ 
- \Big[  &  \langle\hat{n}_{i\lambda,\uparrow}\rangle  \langle \hat{n}_{i\lambda,\downarrow} \rangle  + \Big|\frac{\Delta_{i\lambda}}{U}\Big|^2 \; \Big] , 
\end{split}
\end{equation}
where the self-consistent parameters $\langle \hat{n}_{i\lambda,\sigma} \rangle $ and $\Delta_\lambda=-|U|\langle\hat{c}_{i\lambda\downarrow}\hat{c}_{i\lambda\uparrow}\rangle$ are respectively the occupations and the pairings. $\langle\ldots\rangle $ corresponds to the grand canonical average, here restricted to $T=0$. Because of translation invariance, the cell index is dropped in the self-consistent parameter. Here, we assume a non magnetic ground state, $\langle\hat{n}_{\lambda,\uparrow}\rangle=\langle\hat{n}_{\lambda,\downarrow}\rangle=\langle\hat{n}_{\lambda}\rangle/2$.

The AHM Hamiltonian  given in Eq.~\eqref{H_exact} becomes, 
\begin{eqnarray}
    \hat{H}_{BdG} & = \sum_k 
    \begin{bmatrix}
        \hat{\textbf{C}}_{k \uparrow}^{\dagger} & \hat{\textbf{C}}_{-k \downarrow} \\ 
    \end{bmatrix} 
    \nonumber
    \begin{bmatrix} 
         \hat{h}^{\uparrow}(k)& \hat{\Delta} \\ 
         \hat{\Delta}^{\dagger}& -\hat{h}^{\downarrow *}(-k) \\ 
    \end{bmatrix}
    \begin{bmatrix} 
         \hat{\textbf{C}}_{k \uparrow} \\ 
         \hat{\textbf{C}}_{-k \downarrow}^{\dagger} \\
    \end{bmatrix}
    \\  \nonumber
    & + \sum_{k \lambda} [\hat{h}^{\downarrow}(-k)]_{\lambda\lambda} - \mu -|U| \langle \hat{n}_{i\lambda\uparrow} \rangle  \\ 
    & + \sum_{i \lambda} \Big| \frac{\Delta_\lambda}{|U|} \Big|^2 + |U| \langle \hat{n}_{i\lambda\uparrow} \rangle \langle \hat{n}_{i\lambda\downarrow} \rangle.
\label{H_BdG}
\end{eqnarray}
where we have introduced the spinor $\hat{\textbf{C}}_{k\sigma}^{\dagger} = \Big(
\hat{c}_{k A_1,\sigma}^{\dagger} ,
\ldots , 
\hat{c}_{k A_{N_{CC}},\sigma}^{\dagger},
\hat{c}_{k B_1,\sigma}^{\dagger} ,
\ldots
\hat{c}_{k {B_{N_{CC}}} \sigma}^{\dagger}, \hat{c}_{kC,\sigma}^{\dagger}\Big)^t$ where $c_{k\lambda,\sigma}^{\dagger}$ is the Fourier transform (FT) of $c_{i\lambda,\sigma}^{\dagger}$.  The single particle Hamiltonian $\hat{h}^\sigma (k)=\hat{h}_0(k)-\mu-\hat{V}_{\sigma}$ where $\hat{h}_0$ is the FT of the tight-binding term in Eq.$\,$\eqref{H_exact}. The on-site term $\hat{V}_\sigma=\frac{|U|}{2} \cdot \text{diag}(\langle\hat{n}_{\lambda} \rangle)$ and $\hat{\Delta}=\text{diag}(\Delta_\lambda)$ are $(2 N_{CC}+1)$-diagonal matrices.


Notice that, the constant terms in Eq.$\,$\eqref{H_BdG} have been kept because they play a crucial role for the reliable calculation of the SFW (see below).

\begin{figure*}
    \centering
\includegraphics[scale=0.615]{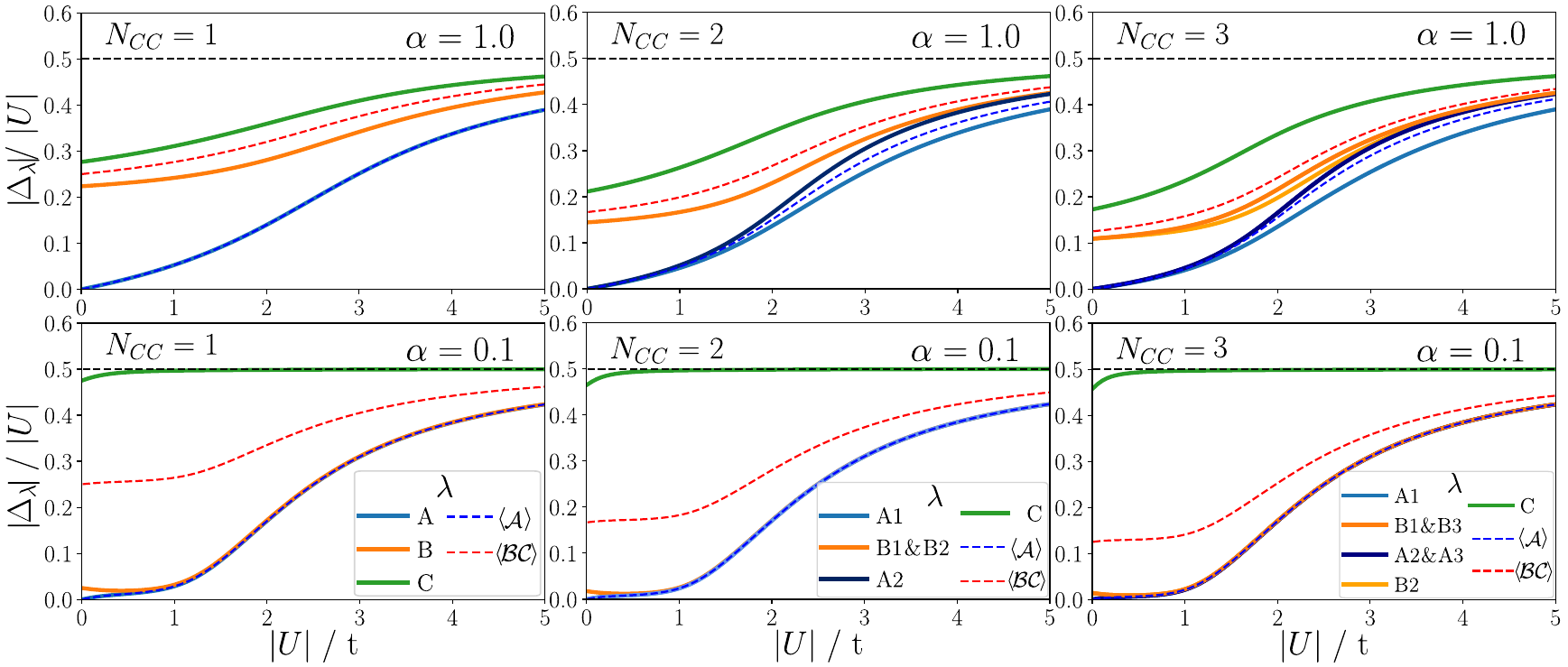}
     \caption{Individual pairings $\Delta_\lambda$ and their average for each sublattice $\mathcal{A}$ and $\mathcal{BC}$, as a function of $|U|/t$. The upper (respectively lower) panels correspond to $\alpha=1.0$ (respectively $\alpha=0.1$) and $N_{CC}=1,2$ and $3$ (from left to right). The filling factor $\nu$ is set to $1$. In each panel, the black dotted line is $(N_\mathcal{BC} \langle \Delta_\mathcal{BC} \rangle - N_\mathcal{A} \langle \Delta_\mathcal{A} \rangle)/|U|$.
     }
 \label{Fig. 4}
\end{figure*}

\section{Results and discussions}
\subsection{Single particle features (U=0)}

First, we recall that the average value of the QM is defined by $\langle g \rangle = \int_{-\frac{\pi}{\tilde{a}}}^{+\frac{\pi}{\tilde{a}}} \frac{dk}{2\pi} g(k)$, where $g(k)=\langle \partial_k \phi^{FB}_k\vert  \partial_k\phi^{FB}_k\rangle - \vert \langle \partial_k \phi^{FB}_k\vert  \phi^{FB}_k\rangle \vert^{2}$, and $\vert  \phi^{FB}_k\rangle$ is the normalized FB eigenstate of the one-particle Hamiltonian which has non-vanishing weight on $\mathcal{BC}$ orbitals only, thus $n_S = N_{CC}+1$. 
In Fig$\,$\ref{Fig. 1}(c) and respectively in Fig$\,$\ref{Fig. 1}(d), the average value of the QM is depicted as a function of $N_{CC}$ for respectively $\alpha = 1.0$ and $\alpha=0.1$. Finding the minimal value of the QM is crucial for what follows since it is the only relevant one \cite{Torma_revisiting}. Keeping in mind that the FB eigenstates have no weight on sublattice $\mathcal{A}$, it is found that the minimal QM is obtained when all B-orbitals are located at the same position, the center of the unit cell: $r_{min}=\Tilde{a}/2$. The full proof is given in the appendix A. For example, for $N_{CC} = 2$ (see Fig$\,$\ref{Fig. 1}(b)) it corresponds to $A_2$'s position. Let us discuss the QM in the case of small out-of-chain hopping, namely $\alpha = 0.1$. We observe that the geometry depicted in Fig$\,$\ref{Fig. 1}(a,b) provides almost the minimal value of $\langle g \rangle$. We remark that both trial functions $aN_{CC}$ or $bN_{CC}^2$ do not provide a satisfactory fit of our data. However, when $N_{CC} \ge 6-7$ the QM is found to scale almost linearly with $N_{CC}$. On the other hand, when $\alpha = 1$, the geometry depicted in Fig$\,$\ref{Fig. 1}(a,b) leads to a strong overestimation of the relevant QM which is amplified as $N_{CC}$ becomes larger and larger. The minimal QM is found to 
scale linearly with $N_{CC}$, more precisely, $\langle g \rangle \simeq 0.47~(N_{CC}-1)$ which is accurate up to values of $N_{CC}$ of about 50. These findings are surprising since one could have naively anticipated a quadratic scaling of the QM with $N_{CC}$ in both cases. Indeed, if we calculate the spreading of the CLS eigenstates $\delta_{CLS}=\langle CLS | \hat{x}_\mu^2 - \langle \hat{x}_\mu \rangle^2 | CLS \rangle$ one finds two distinct regimes: $\delta_{CLS}/a^2 = \frac{1}{4}N^2_{CC}$ when $N_{CC} \alpha^2 \gg 1$ and $\delta_{CLS}/a^2=\frac{1}{12}N^2_{CC}$ when $N_{CC} \alpha^2 \ll 1$. As it will be discussed in the last section the quantity $N_{CC} \alpha^2$ plays an important role. \\
Before we switch on the electron-electron interaction let us briefly discuss the non-interacting case. Figure \ref{Fig. 2} displays the single-particle spectrum and the density of states (DOS)  for $\alpha=1.0$, and $N_{CC} = 1, 2$ and $3$. 
Due to chiral symmetry, the spectrum consists in $2 N_{CC}$ symmetric dispersive bands and at least one FB at $E=0$. Here, it appears that for any $N_{CC}$ the spectrum exhibits a single FB. Numerically it is found that the gap between the FB and the first dispersive band decreases as $\delta \simeq t/N_{CC} + o(1/N_{CC}^2)$. We remark that the DOS possesses $4 N_{CC}$ van-Hove singularities.
\\

\subsection{Pairings and quasi-particles gap}
Let us define two regions for the filling factor $\nu = \sum_{\lambda,\sigma} n_{\lambda,\sigma} / N_{orb} \in [0,2]$: (i) the FB region which corresponds to $\nu \in [1-\frac{1}{N_{orb}},1 +\frac{1}{N_{orb}}]$, and (ii) the DB region which corresponds to any other filling. Notice that chiral symmetry arising from bipartite nature of the lattice implies a symmetry of the observables with respect to half-filling ($\nu=1$). \\
Fig$\,$\ref{Fig. 3} depicts the pairings for each orbital as a function of $\nu$. For $\alpha=1.0$, as $N_{CC}$ increases, multiple domes and peaks appear in the DB region due to new dispersive bands in the spectrum, simultaneously the pairing amplitudes are rising. While the FB region shrinks, the pairings at half-filling decrease in a moderate way. For low carrier concentration, $\nu \le 0.1$ or $\nu \ge 1.9$, we find that the pairing in the one-dimensional chain (ODC) which corresponds to $\alpha = 0$ agrees well with the average value of the pairings along the chain.
In contrast, for $\alpha = 0.1$, the changes are drastic, the multi-peaks and domes have disappeared. It can be seen that the increase of $N_{CC}$ has no qualitative and weak quantitative impact on the pairings. Indeed, the $C-$orbital being weakly coupled to the chain, makes $A$-type and $B$-type orbitals quasi-equivalent leading to quasi-uniform pairings close to that of the ODC.
The only trivial effect of $N_{CC}$ is to shrink the FB region. Furthermore, the $C$-orbital being quasi-disconnected, $\Delta_C$ has a tiny value in DB region, and almost reaches its maximum value in the FB region where $\Delta_C \simeq \frac{|U|}{2}$ at half-filling. It is worth noticing that for $\alpha=1$ and $|U| \gg t$ the  multi-peak structure observed previously disappears, the situation become similar to that of $\alpha=0.1$.

Focusing now on the half-filled case ($\nu=1$), Fig.$\,$\ref{Fig. 4} displays the individual pairings $\Delta_\lambda/|U|$ and their average as a function of $|U|$, for different values of $\alpha$ and $N_{CC}$. $\langle \Delta_\Lambda \rangle$ means average pairing for sublattice $\Lambda=\mathcal{A}$ and $\mathcal{BC}$.
In weak coupling regime and for any $N_{CC}$, the FB eigenstates having a non-zero weight on sublattice $\mathcal{BC}$ only, implies that the pairings associated with this sublattice scale linearly with $|U|$ while they are quadratic on sublattice $\mathcal{A}$ \cite{Gap_linear_1990, Gap_linear_1994,Peotta_Nature}. 
It can be seen that the dilution of the FB-states leads to the suppression of the pairing amplitudes. More precisely, for $|U|\le t$, the average pairing on $\mathcal{BC}$ is found to decay as $\frac{|U|}{2(1+N_{CC})}$ for both values of $\alpha$. Thus, it smoothly vanishes as 
$N_{CC}$ increases.
For $\alpha=1.0$, as $N_{CC}$ grows, one systematically finds that the individual pairings on B-orbitals are larger than $\langle \Delta_\mathcal{A} \rangle$  and smaller than $\langle \Delta_\mathcal{BC} \rangle > \langle \Delta_\mathcal{A} \rangle$.
In addition, as already seen in Fig.\ref{Fig. 3}, when $\alpha=0.1$, all $A$s and $B$s sites become quasi-equivalent while $\Delta_C \simeq \frac{|U|}{2}$ due to weak coupling to the chain. Lastly, we have checked that when $N_{CC} \gg 1$, the pairings on sublattice $\mathcal{A}$ and the QP gap converge towards that of the isolated chain $\Delta^{ODC}$.
Finally, note that the pairing sum rule $N_\mathcal{BC} \langle \Delta_\mathcal{BC} \rangle - N_\mathcal{A} \langle \Delta_\mathcal{A} \rangle = \frac{|U|}{2}$ is illustrated as well in Fig.\ref{Fig. 4} \cite{Thumin_EPL_2023,Bouzerar_SciPost_2024}.

\begin{figure}[h!]
    \centering
    \includegraphics[scale=0.55]{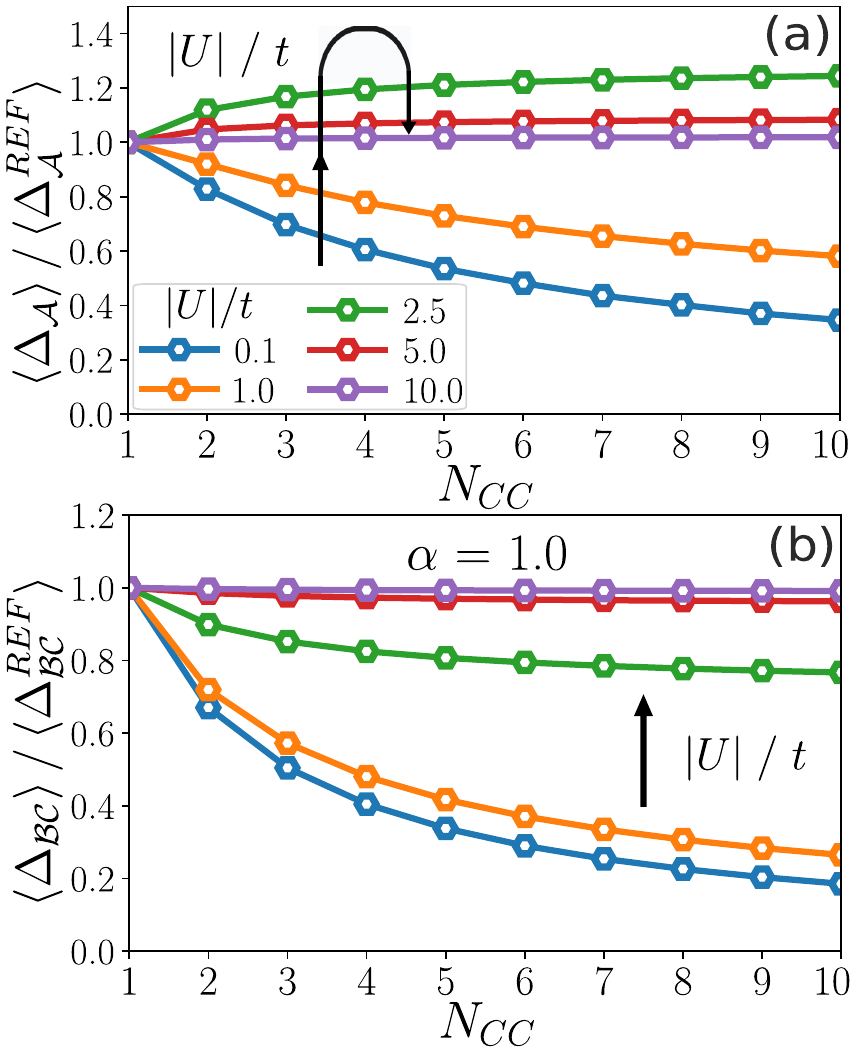} 
    \caption{Average values of the pairings $\langle \Delta_\Lambda \rangle$ as a function of $N_{CC}$ where $\Lambda=\mathcal{A}$ (panel (a)) and $\Lambda=\mathcal{BC}$ (panel (b)).
    The pairing are rescaled by their value at $N_{CC}=1$ ($\langle \Delta_\Lambda^{REF} \rangle$). Different values of $|U|/t$ are considered and $\alpha = 1.0$. The carrier density is set to $1$ (half-filling).}
    \label{Fig. 5}
\end{figure}

Fig.$\,$\ref{Fig. 5} (a) and (b) show the average pairing on both sublattice as a function of $N_{CC}$ for $\alpha = 1.0$ and different values of $|U|$. First, $\langle \Delta_\mathcal{A} \rangle$ displays two different behaviours depending on $|U|$. In the weak coupling regime, it is found that it decreases as $N_{CC}$ increases, while, when $|U| \gtrsim 1.5 t$ it increases. In contrast, one observes 
in Fig.\ref{Fig. 5}\textbf{(b)} that
$\langle \Delta_\mathcal{BC} \rangle$ decreases regardless the interaction strength, the reduction being significantly stronger in the case of small values of $|U|$. For $|U| \gg t$, for any value of $N_{CC}$ it can be seen for both sublattices that $\langle \Delta_\Lambda \rangle = \langle \Delta_\Lambda^{REF} \rangle$, where $\Lambda = \mathcal{BC}$ or $\mathcal{A}$.
To summarize our findings, in the weak coupling regime, the dilution process tends to reduce the average binding energy of Cooper pairs on both sublattices.  \\

\begin{figure*}
    \centering
\includegraphics[scale=0.43]{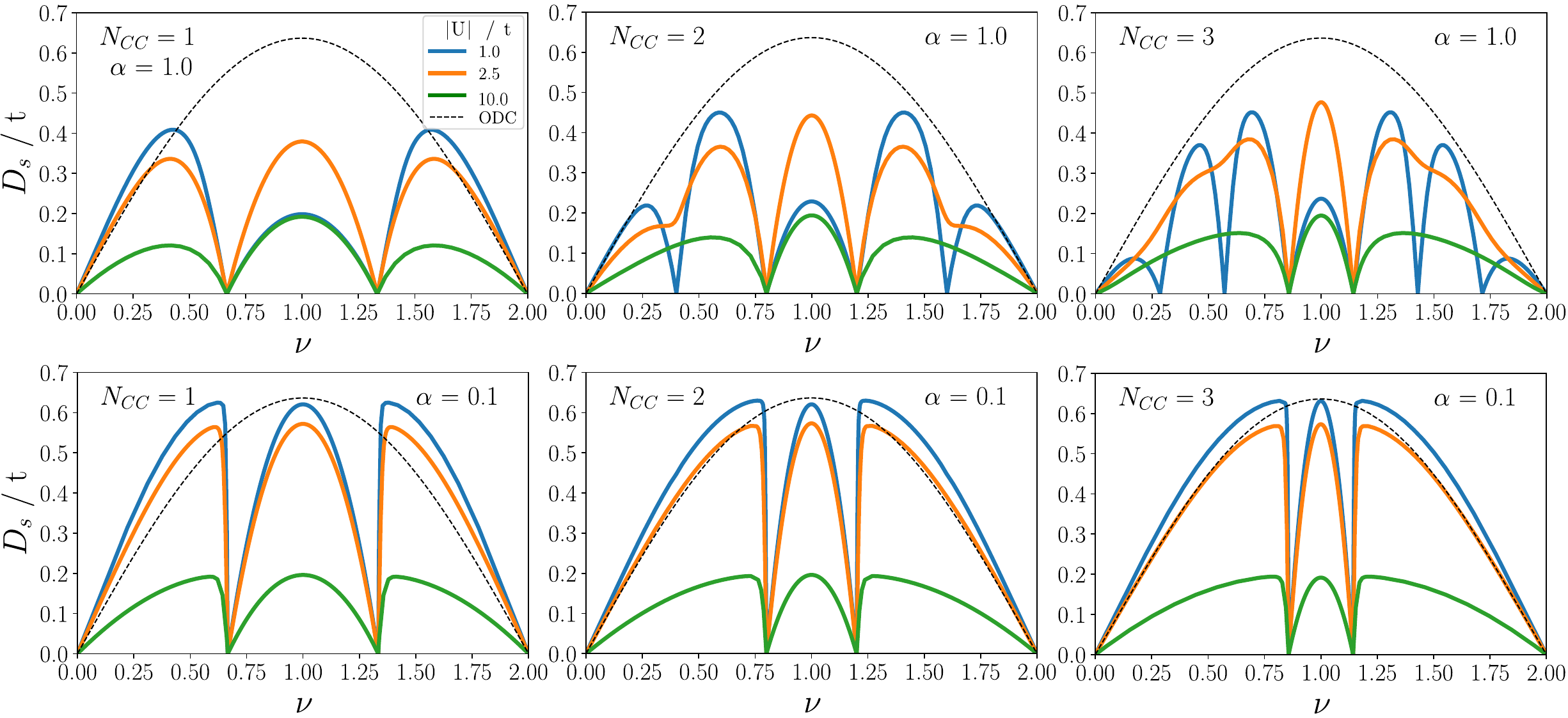} 
     \caption{Superfluid weight $D_s$, as a function of the filling factor $\nu$. 
     The upper panels correspond to $\alpha=1.0$, and $N_{CC}=1, 2$ and $3$ (left to right). 
     The lower panels correspond to $\alpha=0.1$ and the same values of $N_{CC}$. Blue, orange, and green curves  represent respectively $|U|/t=1, 2,$ and $10$. For comparison $D_s$ in the non-interacting one-dimensional chain ($\alpha=0$ and $|U|=0$) is depicted as well ('ODC').  
     }
\label{Fig. 6}
\end{figure*}

\subsection{The superfluid weight}
The SFW $D_s$ \cite{Khon_Ds, Shastry_Sutherland_Ds, Scalapino_Ds} is the superconducting order parameter that is directly related to London's penetration $\lambda_L \propto 1/\sqrt{D_s}$. The SFW is carefully defined as,
\begin{equation}
D_s=\frac{1}{N_c \tilde{a}}\frac{d^2\Omega(q)}{dq^2} \Big|_{q=0} .
\label{def_Ds}
\end{equation}
where $N_c$ is the number of unit cells in the lattice, $\Omega (q)$ being the grand-potential and $q$ mimics the effect of a vector potential introduced by a standard Peierls substitution $t^{\lambda\eta}_{ij} \longrightarrow t^{\lambda\eta}_{ij} e^{iq(x_{i\lambda}-x_{j\eta})}$. It is important to notice that the derivative is total. It includes both the explicit and the implicit (through the self-consistent parameters) $q$-dependence, the latter being essential for reliable calculations of the SFW \cite{Torma_revisiting}.

Fig.$\,$\ref{Fig. 6} depicts $D_s$ as a function of $\nu$ for $\alpha = 0.1$ and $1.0$, and $N_{CC} = 1, 2,$ and $3$. First, for $\alpha = 1.0$, the increase of $N_{CC}$ leads to a multi-dome structure as seen already in the pairings, with special fillings for which $D_s$ vanishes and corresponding to fully filled one-particle bands. 
In the DB region and away from these specific points, as $|U|$ increases the SFW reduces. For any value of $N_{CC}$, in the strong coupling regime the multi-dome structure disappears.
Regarding the FB superconductivity, we observe a moderate increase as $|U|$ and $N_{CC}$ increase, besides the trivial shrink of the FB region.
A more detailed analysis of the specific half-filled case is discussed in next section.
On the other hand, for $\alpha=0.1$, the multi-dome structure is absent for any value of the interaction strength as observed in Fig.\ref{Fig. 3} for the pairings. As $N_{CC}$ increases, the decimation has a relatively weak effect for these values of $|U|$. Finally, as expected, in the very dilute regime, $D_s(\nu)$ converges towards the single chain limit which is given by $D^{ODC}_s(\nu)=\frac{2a}{\pi}\sin(\pi\nu/2)$.
Based on the scaling of $\langle g \rangle$ plotted in Fig.\ref{Fig. 1} (c) and (d), these findings are unexpected and surprising, since we could have anticipated, at least for the smallest value $|U|=1$ a significant increase of the SFW. We shed light on these findings in the next section.

\begin{figure}[h!]
    \centering
    \includegraphics[scale=0.4]{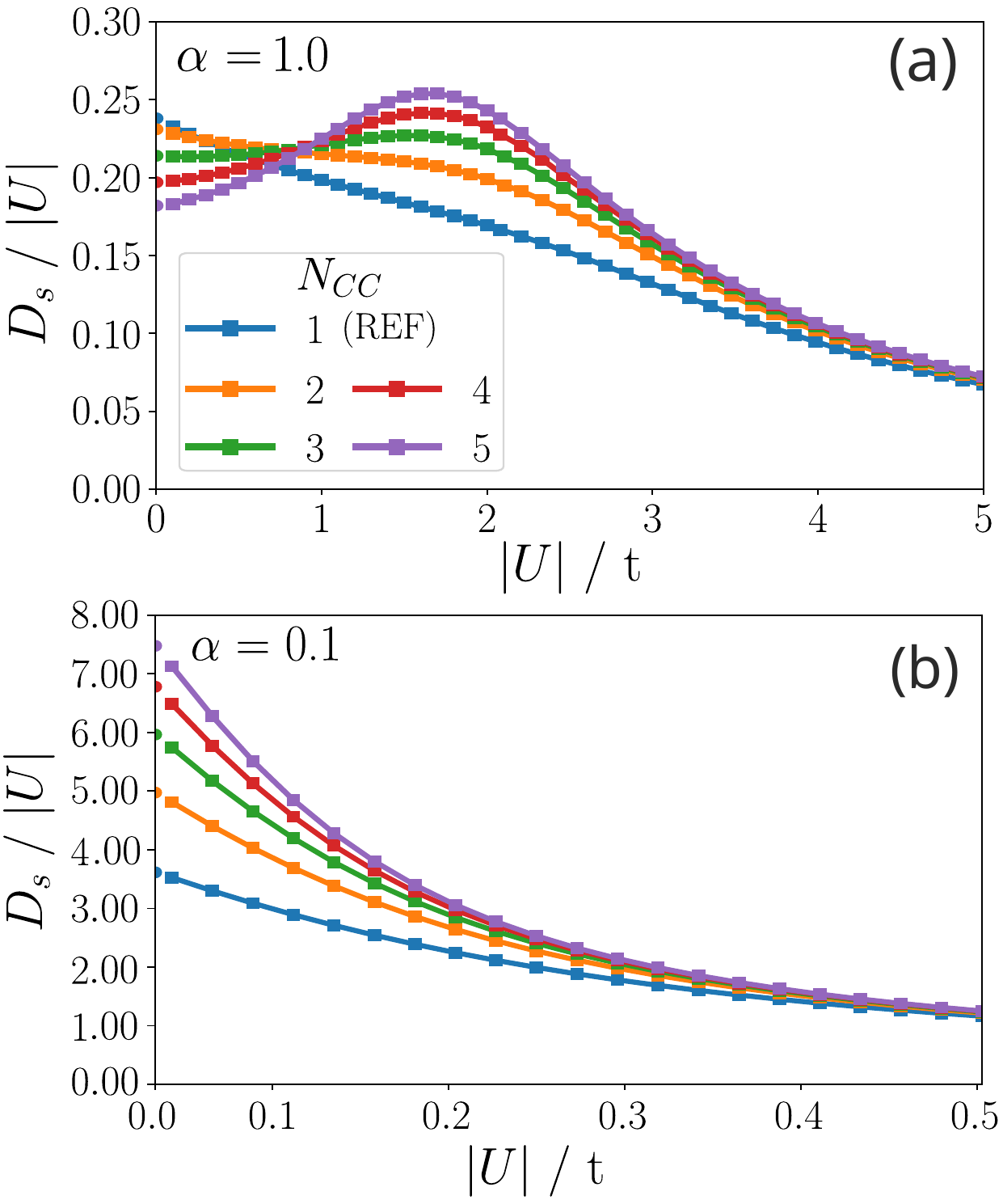} 
    \caption{$D_s/|U|$ in the half-filled dilute stub chain as a function of $|U|$ for $N_{CC} = 1, 2, 3, 4$ and $5$, and respectively $\alpha=1.0$ in panel \textbf{(a)}, and $\alpha=0.1$ in panel \textbf{(b)}. Circular dots correspond to extrapolated values at $|U|=0$. 
    }
    \label{Fig. 7}
\end{figure}
\subsection{Effects of flat band states dilution on the superfluid weight}

From now on, we focus essentially on the half-filled case ($\nu=1$). In Fig.\ref{Fig. 7}, $D_s/|U|$ is displayed as a function of $|U|/t$ for $N_{CC} = 1,\hdots,5$. First, we propose to discuss the case $\alpha = 1.0$ which is depicted in Fig.\ref{Fig. 7}(a). One can observe two distinct regimes. In the weak coupling regime, $D_s$ reduces as $N_{CC}$ increases, which is the opposite of what could have been expected. We will come back to this confusing point later.
On the other hand, for $|U|/t \ge 1$, the SFW becomes larger and larger as $N_{CC}$ increases. Here, the maximal augmentation is reached for $|U|/t \simeq 1.8$. 
Finally, for $|U|\gg t$ it is found that the dilution has essentially no impact. More precisely one finds that $D_s \approx \frac{2t^2}{|U|} a$ which is demonstrated analytically in the Appendix B.
We now consider the case of the weak hopping between the C-orbitals and the chain ($\alpha=0.1$). Our findings are depicted in Fig.\ref{Fig. 7}(b) where
a single regime is observed: in the weak coupling regime $D_s$ is boosted by the dilution of the FB eigenstates. In contrast, when $|U|/t \ge 5\alpha$ it is observed the dilution has no impact. We observe that $D_s \rightarrow \frac{2t^2}{|U|}$ as found for $\alpha = 1$ (see appendix B). In the weak coupling regime, it is surprising that the dilution leads to an increase of the SFW only for the smallest values of $\alpha$.

\begin{figure}[h!]
    \centering
    \includegraphics[scale=0.55]{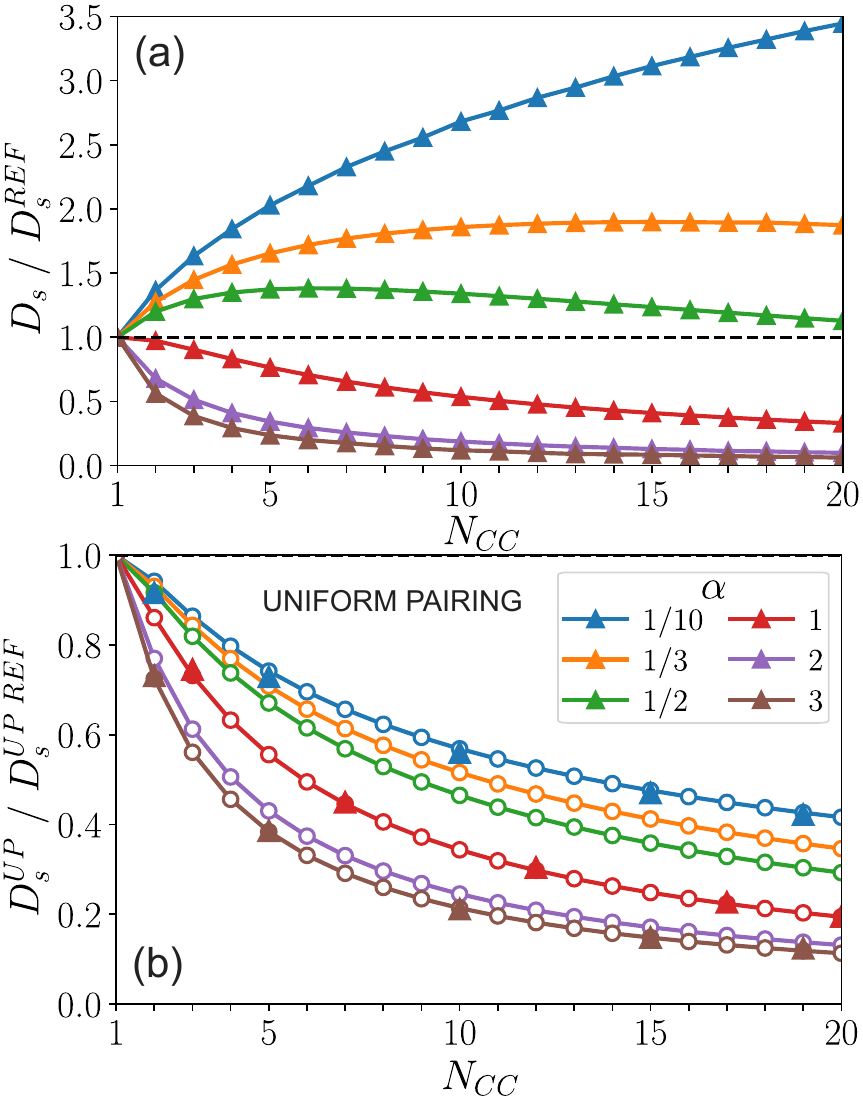} 
    \caption{Superfluid weight in the weak coupling regime, rescaled by its value at $N_{CC}=1$ (REF), as a function of $N_{CC}$. The system is half-filled and $\alpha$ ranges from $0.1$ to $3.0$. In panel \textbf{(a)}, the complete self-consistent calculation of the SFW. In panel \textbf{(b)}, the SFW obtained by forcing the uniform pairing condition. Empty circles correspond to the analytical formula given in Eq.$\,$\eqref{Ds_analytical}, and the filled triangles to the numerical calculations (in both panels).
    }
    \label{Fig. 8}
\end{figure}

The previous paragraph seems to point out the importance of the value of the out-of-chain hopping. To complete our analysis, we plot in 
Fig.\ref{Fig. 8}, the SFW in the weak coupling regime as a function of $N_{CC}$ for values of $\alpha$ ranging from 0.1 to 3.
In 
Fig.\ref{Fig. 8} (a) one clearly observe two different type of behaviour. For $\alpha \ge 1$ the dilution systematically suppresses the SFW with respect to its value at $N_{CC}=1$. The reduction becomes more pronounced as $\alpha$ increases. In contrast, when $\alpha \le 1$ the effect of the dilution is the opposite: $D_{S}$ is non-monotonous, it increases till it reaches a maximum then it slowly decay. The maximum is clearly visible for $\alpha=0.5$, and it is relatively broad for $\alpha=0.3$. On the other hand, for $\alpha=0.1$ the maximum has not been reached yet for the considered values of $N_{CC}$.
A careful analysis reveals that the crucial parameter that controls the existence of a maximum is $N_{CC} \alpha^2$. More precisely it is found that the maximum occurs when $N_{CC} \alpha^2 \approx 1.5$. Below this value the SFW increases and above it decays. This explains why no maximum is observed when $\alpha \ge 1$.
{\it How can we understand the emergence of such a parameter and this specific value of 1.5?}
There is no simple answer, however it is useful to return to the CLS eigenstates.
For a given $N_{CC}$ the total weight on B-orbitals is exactly $N_{CC}\alpha^2$ while it is 2 for the C-orbitals. Thus, when $N_{CC} \alpha^2 \ge 2$ (respectively $\le 2$), the dominant weight is on B-orbitals (respectively on C-orbitals). The overlap between two CLS that share one C-orbital is $\frac{1}{N_{CC}\alpha^2+2}$. When $\alpha \ge  1$ this ratio decays rapidly as $N_{CC}$ increases which is detrimental for the transport along the chain. On the other hand, when $\alpha \ll 1$ this ratio remains almost constant ($\approx 1/2$) and decays significantly only when $N_{CC} \ge 2/\alpha^2$ which for $\alpha=0.1$ is 200. Within our simple explanation, in the first stage one would expect $D_s$ to be almost constant. The fact that an increase is found in our numerical calculations cannot be captured with these heuristic arguments. We conclude that for $\alpha \le 1$ the dilution of the FB eigenstates leads to a boost of the SFW. \\

In the literature, one often assumes the UP condition that considerably simplifies the calculations and allows analytical results.
In this last paragraph, we propose to discuss our findings by comparing them to what we would obtain if we force the UP hypothesis. We consider the case of weak coupling regime only.
We recall that, in half-filled FB system when $|U|$ is smaller than the one-particle gap the SFW reads \cite{Peotta_Nature},
\begin{eqnarray}
  D^{UP}_s = \frac{ 2\langle g \rangle}{n_S \Tilde{a}} |U|.  
  \label{Ds_analytical}
\end{eqnarray}
In Fig.\ref{Fig. 8}(b), $D^{UP}_s$ is plotted as a function of $N_{CC}$. First, one observes that the numerical calculations are in perfect agreement with the analytical formula.
In addition, it can be seen that the UP condition leads to a monotonic and systematic decay of the SFW as $N_{CC}$ increases. The reduction being amplified in the case of a strong out-of-chain hopping. This contrasts with the full (no UP condition) calculations plotted in Fig.\ref{Fig. 8}(a). Furthermore, in the case where $\alpha \ge 1$ the ratio $D_s/D^{REF}_s$ is found to decay more rapidly in the complete self-consistent calculation.
Finally, we emphasize that the numerical value of $D^{UP}_s$, given in Eq.~\ref{def_Ds} were obtained for the geometry that minimizes the QM. Surprisingly it is revealed that $D^{UP}_s$ is geometry-dependent which indicates that forcing the UP condition may lead to unreliable values of the SFW. Thus the UP approximation should be used with caution especially in nano-structured systems and in disordered lattices. 

\section{Conclusion}

In this work we have addressed in details the impact of the dilution of FB states on the minimal quantum metric, on the pairing distribution and on the SFW. The connection between the SFW and the quantum metric is revealed to be more complex than predicted by recent theories. Furthermore, for weak out-of-chain hopping, we have established that the SFW can be boosted by the decimation of the FB eigenstates. Whereas when this hopping is comparable or larger to that along the chain the SFW turns out to be suppressed. Additionally, it is revealed that the widely used uniform paring hypothesis appears to be qualitatively incorrect in such dilute systems.
In the quest for unconventional high critical temperature superconducting materials 
these findings could prove very useful. We could think of well-chosen atoms deposited in a controlled manner on the surface of a specific quasi two dimensional material which could give rise to weakly dispersive bands with a large quantum metric. To target the best candidates, two dimensional materials and ad-atoms, one could rely on first-principle studies which offer great advantages and flexibility as well. This type of strategy has proved to be powerful in addressing the issue of ferromagnetism in diluted magnetic semiconductors.

\section*{Appendix A: Flat band states and minimal quantum metric}
\numberwithin{equation}{section}
\setcounter{equation}{0}
\renewcommand{\theequation}{A.\arabic{equation}}
It is well known that the quantum metric (QM) is not uniquely defined, it depends on the position of the orbitals inside the unit cell \cite{Torma_revisiting}. However, the only physically relevant one, connected to the superfluid weight, is its minimal value $\langle g_{min} \rangle$.\\
In this appendix our purpose is to demonstrate that in the dilute stub lattice, for a given $N_{CC}$, the minimal QM $g_{min}$ is obtained when $B_i$ atoms where $i$ = 1, 2,..., $N_{CC}$ are all placed at the center of the unit cell. For instance, for the $N_{CC} = 2$, as depicted in Fig.\ref{Fig. 1}, $g_{min}$ is obtained when $B_{1}$ and $B_{2}$ coincide with the location of $A_{2}$.
First, because the flat band (FB) eigenstates have a finite weight only on C and $B_i$ sites, the position of the atoms of type A is irrelevant. 
Let us define, $\delta_i$, where the index $i$ = $1$, $2$,$ ...$, and $N_{CC}$, the position (coordinate along the chain) of the atoms of type B inside the unit cell. 

As a first step, using the expression of the CLS state, one can show that the $k$-dependent normalized FB eigenstate of the single particle Hamiltonian reads,
\begin{equation}
 \vert \phi^{FB}_k \rangle = \frac{1}{D(k)}     
       \begin{pmatrix}
         1-(-1)^{N_{CC}}e^{ik\Tilde{a}} \\ 
         -\alpha e^{ik\delta_1} \\
         +\alpha e^{ik\delta_2} \\
         . \\
         . \\
         . \\
         (-1)^{N_{CC}}\alpha e^{ik\delta_{N_{CC}}}\\
    \end{pmatrix}
    ,
\end{equation}
where the FB state is expressed in the basis $\Big (\hat{c}_{kC,\sigma}^{\dagger},
\hat{c}_{k B_1,\sigma}^{\dagger},\ldots,\hat{c}_{k {B_{N_{CC}}},\sigma}^{\dagger}\Big)$ and  $D(k)=\sqrt{u+v\cos(k\Tilde{a})}$ where $u = N_{CC}\alpha^2 +2$ and $v = -2(-1)^{N_{CC}}$ and we recall that $\Tilde{a}=N_{CC}\cdot a$.\\

In order to calculate the quantum metric, we start from its definition,
\begin{equation}
g(k)=\langle\partial_{k}\phi^{FB}_k \vert \partial_{k} \phi^{FB}_k \rangle -\vert \langle \partial_{k} \phi^{FB}_k \vert \phi^{FB}_k \rangle\vert^{2}.
\end{equation}
After some lengthy but straightforward calculation we end up with,
\begin{equation}
\begin{split}
\langle g \rangle = 
&\int \frac{\alpha^2}{D(k)^2} \Big[ \delta_1^2+\delta_2^2+....+\delta_{N_{CC}}^2 \Big] 
+ \int h(k)  \\ 
- &  \int \frac{1}{D(k)^4} \Big[ 
\alpha^2(\delta_1+\delta_2+....+\delta_{N_{CC}}) + 2\Tilde{a}
f(k) \Big]^2, \\
\end{split}
\label{qmetric}
\end{equation}
where $\int (..) = \frac{1}{2\pi}\int_{-\pi/\Tilde{a}}^{+\pi/\Tilde{a}} dk(..)$, the function $f(k)=\sin^2(k\Tilde{a}/2)$ (respectively $\cos^2(k\Tilde{a}/2)$) if $N_{CC}$ is even (respectively odd).
For the present purpose, it is unnecessary to give explicitly the expression of $h(k)$ which does not depend on $\delta_i$'s.\\
To find the optimal position of the atoms of type B, one has to solve for $i=1,2,...,N_{CC}$, the equation,
\begin{eqnarray}
\frac{\partial \langle g \rangle }{\partial \delta_i} = 0,
\end{eqnarray}
which leads to,
\begin{eqnarray}
\delta_i = \frac{\alpha^2 I_2 S + 2\Tilde{a}I_3}{I_1},
\end{eqnarray}
where $S=\Sigma_{i} \delta_i$, and the integrals are respectively $I_1=\int \frac{1}{D(k)^2}$, $I_2=\int \frac{1}{D(k)^4}$ and $I_3=\int \frac{f(k)}{D(k)^4}$.
This can be simplified and finally gives,
\begin{eqnarray}
\delta_i = \frac{S}{N_{CC}}=\frac{2\Tilde{a}I_3}{I_1-wI_2},
\label{eqdeltamin}
\end{eqnarray}
where we have introduced $w = N_{CC}\alpha^2$.\\
This equation means that the minimal value of the quantum metric corresponds to the situation where the atoms $B_i$'s are all located at the same $\alpha$-dependent position.\\
Interestingly, these three integrals can be calculated analytically and one obtains,
\begin{eqnarray}
I_1 = \frac{1}{(w^2+4w)^{1/2}}; \, \, I_2 = \frac{2+w}{(w^2+4w)^{3/2}}.
\end{eqnarray}
The last integral $I_3$ depends on the parity of $N_{CC}$,
\begin{eqnarray}
I_3= \frac{s(w)}{2(w^2+4w)^{3/2}}. 
\end{eqnarray}
where we find $s(w)=w$ for both $N_{CC}$ even and odd. This allows great simplifications, and leads to,
\begin{eqnarray}
\delta_i = \Tilde{a}\frac{s(w)}{2w}=\frac{1}{2}\Tilde{a}
\end{eqnarray}
Thus, for a given $\alpha$ the minimal quantum metric corresponds to B-atoms located at the center of the supercell. 
Finally, we notice that, using (i) the three integrals $I_1$, $I_2$ and $I_3$, (ii) the fact that $\delta_i=\frac{1}{2}\Tilde{a}$ and last (iii) the exact expression of $h(k)$ (not given here), from Eq.\eqref{qmetric} one will get the exact analytical formula for the minimal QM as a function of $\alpha$ and $N_{CC}$.

\section*{Appendix B: Superfluid weight in the strong coupling regime}
\numberwithin{equation}{section}
\setcounter{equation}{0}
\renewcommand{\theequation}{B.\arabic{equation}}

In this appendix, we demonstrate in the case of the half-filled dilute stub lattice and in the strong coupling regime ($|U|\gg t$) that the superfluid weight (SFW) $D_s = \frac{2t^2}{|U|}a$. This implies that $D_s$ is independent of both $\alpha$ and $N_{CC}$.

Following the procedure of Refs.\cite{Dice_chinois, Thumin_PRB_2023}, at $T=0~K$ the SFW can be written, 
\begin{equation}
    D_s = \frac{2}{N_c\Tilde{a}} \sum_{k,mn} \frac{|\bra{\Psi_{nk}^-}\hat{V}\ket{\Psi_{mk}^+}|^2 -|\bra{\Psi_{nk}^-}\hat{\Gamma}\hat{V}\ket{\Psi_{mk}^+}|^2}{E_{nk}^- - E_{mk}^+},
    \label{Ds_V}
\end{equation}
where $\ket{\Psi_{mk}^\pm}$ are the quasi-particle eigenstates of the BdG Hamiltonian with positive $(+)$ and negative $(-)$ energy. The band indices $m$ and $n$ vary from 1 to $N_{orb} = 2N_{CC}+1$. $N_c$ denotes the number of unit cells and $\Tilde{a} = N_{CC}\cdot a$ is the size of the unit cell in the dilute stub lattice. The matrices $\hat{\Gamma}=\text{diag}(\mathbb{\hat{I}}_{N_{orb}\times N_{orb}}, -\mathbb{\hat{I}}_{N_{orb}\times N_{orb}})$ and $\hat{V} = \text{diag}(\hat{v}_0,\,\hat{v}_0)$, $\hat{v}_0$ being the velocity operator,
\begin{eqnarray}
  \hat{v}_0(k) = \frac{\partial\hat{h}_0(k)}{\partial k},  
\end{eqnarray}
where $\hat{h}_0(k)$ is the single particle Hamiltonian.\\

In the natural basis
$\mathcal{B_{N_{CC}}}=\Big(\vert kC\sigma \rangle ,\,
\vert kA_1\sigma \rangle ,\,
\vert kB_1\sigma \rangle ,\,
\ldots, 
\vert kA_{N_{CC}}\sigma \rangle ,\,
\vert kB_{N_{CC}}\sigma \rangle
\Big)$ the Hamiltonian $\hat{h}_0(k)$ has a trigonal form and reads, 
\begin{eqnarray}
\hat{h}_0(k)=-t
   \begin{bmatrix} 
         0&\alpha&0&.&.&.&.&0\\ 
         \alpha&0&\beta*&.&.&.&.&\beta \\
         0&\beta&0&\beta^*&.&.&.&0 \\
         0&0&\beta&0&\beta^*&.&.&0\\
        0&.&.&.&.&.&.&0\\  
        0&.&.&.&.&.&.&0\\   
        0&.&.&.&.&\beta&0&\beta^*\\  
      0&\beta^*&.&.&.&.&\beta&0\\
   \end{bmatrix}
   \label{hok}
\end{eqnarray}
where $\beta= e^{ika/2}$.\\
In the strong coupling regime ($|U|\gg t$) and for the half-filled lattice, the pairings are uniform, $\Delta_\lambda = \Delta = \frac{|U|}{2}$, and the quasi-particle energies reduce $E_n^\pm = \pm|\Delta| + o(\frac{t^2}{|U|})$, the corresponding eigenstates are given by,
\begin{equation}
    \ket{\Psi^+_{nk}}= \frac{1}{\sqrt{2}}
    \begin{bmatrix}
           \ket{\phi_{nk}} \\
           \ket{\phi_{nk}} \\
    \end{bmatrix}
    \,
    \ket{\Psi^-_{nk}}=  \frac{1}{\sqrt{2}}
    \begin{bmatrix}
           +\ket{\phi_{nk}}\\
           -\ket{\phi_{nk}} \\
         \end{bmatrix},
\label{phi-def}
\end{equation}
where $\ket{\phi_{nk}}$'s are the one-particle eigenstates of $\hat{h}_0(k)$.

One can show that,
\begin{equation}
\begin{split} \bra{\Psi_{nk}^-}\hat{\Gamma}\hat{V}\ket{\Psi_{mk}^+} &= - \bra{\phi_{nk}}\hat{v}_0\ket{\phi_{mk}}  \\ 
    \bra{\Psi_{nk}^-}\hat{V}\ket{\Psi_{mk}^+} &= 0.
\end{split}
\end{equation}

Hence, Eq.\ref{Ds_V} can be simplified and gives,
\begin{equation}
    D_s = \frac{2}{|U|N_c\Tilde{a}} \sum_{k,nm} |\bra{\phi_{nk}}\hat{v}_0\ket{\phi_{mk}}|^2=\frac{2}{|U|N_c\Tilde{a}} \text{Tr}[\hat{v}_0^2] .
\end{equation}
Given the tridiagonal form of $\hat{h}_0(k)$ as given in Eq.(\ref{hok}), one can easily deduce the diagonal matrix elements of $\hat{v}_0^2$. Indeed, straightforwardly one finds $\langle v_i \vert \hat{v}_0^2 \vert v_i \rangle = \frac{1}{2}(ta)^2$, where $\vert v_i \rangle$ is either $\vert kA_i\sigma \rangle $ or $\vert kB_i\sigma \rangle $, with $i=1,...,N_{CC}$.
Thus, we end up with a simple expression of the SFW in the strong coupling regime that does not depend on $\alpha$ or $N_{CC}$,
\begin{equation}
D_s = \frac{2t^2}{|U|}a
\end{equation}


\end{document}